\documentclass[aps,preprint]{revtex4}

\usepackage[dvips]{graphicx}

\newcommand{\ch}{{\cal H}}
\newcommand{\ce}{{\cal E}}
\newcommand{\cpt}{PT}

\begin{document}

\date{\today}

\title{Clustering of exceptional points and dynamical phase transitions}
\author{Hichem Eleuch$^{1,2}$\footnote{email: hichemeleuch@yahoo.fr} and 
Ingrid Rotter$^{3}$\footnote{email: rotter@pks.mpg.de}}

\address{
$^1$Department of Physics, McGill University, Montreal, Canada H3A 2T8}
\address{
$^{2}$ Department of Physics, Universit\'{e} de Montr\'{e}al, Montreal, QC,  H3T 1J4, Canada
}
\address{
$^3$Max Planck Institute for the Physics of Complex Systems,
D-01187 Dresden, Germany }

\begin{abstract}
  
The eigenvalues of a non-Hermitian Hamilton operator are complex and provide
not only the energies but also the lifetimes of the states of the system. 
They show a non-analytical behavior at singular (exceptional) 
points (EPs). The eigenfunctions are biorthogonal,
in contrast to the orthogonal eigenfunctions of a Hermitian operator. 
A quantitative measure for the ratio between biorthogonality and 
orthogonality is the phase rigidity of the wavefunctions. 
At and near an EP, the phase rigidity takes its minimum value.
The lifetimes of two nearby 
eigenstates of a quantum system bifurcate under the influence of an EP. 
When the parameters are tuned to the point of maximum width bifurcation, 
the phase rigidity suddenly increases up to its maximum
value. This  means that the eigenfunctions become almost orthogonal at 
this point. This unexpected result is very robust as shown by
numerical results for different classes of systems.
Physically, it  causes an irreversible stabilization of the system by 
creating local structures that can be described well by a Hermitian 
Hamilton operator. Interesting 
non-trivial features of open quantum systems appear in the parameter 
range in which a clustering of EPs causes a dynamical phase transition.

\end{abstract}

\pacs{\bf }
\maketitle

\section{Introduction}
\label{intr}

In its simplest form, the wavefunction of a 
quantum many-particle fermionic system can be written
as a Slater determinant that satisfies the anti-symmetry requirements 
and consequently the Pauli principle.  However, some corrections
in the wavefunctions are usually necessary which are caused by their 
possible mixing. In nuclei, for example, corrections arise  
from the residual interaction between the particles of the system, 
with the consequence that the wavefunctions of all the states are 
strongly mixed.
Another source is the interaction between system and environment 
into which the system is embedded. In this case, the states of
the system interact via the common environment due to which  
their wavefunctions may be modified. 

The natural environment of a  localized quantum mechanical system 
is the  extended continuum of scattering wavefunctions in which
the system is embedded. This environment can be 
changed by means of external forces, however it can never be deleted. 
It causes some communication between distant levels, for details see 
the recent review \cite{ropp} in which theoretical results are 
confronted with experimental results. The theoretical results
are obtained by using a non-Hermitian Hamilton operator $\ch$
for the description of the system, 
which contains explicitly the interaction between system and 
environment \cite{top}. The eigenvalues and eigenfunctions 
of $\ch$ differ essentially from those of a Hermitian Hamilton 
operator\,: the eigenfunctions are biorthogonal and the eigenvalues
may show, as function of a certain parameter, deviations from
Fermi's golden rule. The differences between the
eigenvalues and eigenfunctions of $\ch$ and those of a Hermitian 
operator $H$ appear, above all, at and near to 
singular points at which two eigenvalues of $\ch$ coalesce and 
the corresponding eigenfunctions
differ from one another only by a phase. These singular points
do not have any equivalent in the mathematics based on Hermitian 
Hamiltonians since the different eigenfunctions of $H$ are 
orthogonal (in contrast to those of $\ch$, which are biorthogonal). 
These singular points are called usually exceptional points 
(EPs), according to \cite{kato}. 

The role of EPs in physical systems is considered in many papers 
during last years, e.g. \cite{mondragon,pastawski,mois2,atabek,jolicard}, 
see also the review \cite{top} and the book \cite{moiseyev}. In the 
present paper, we are interested in the relation of these singular
points to phase transitions occurring in open quantum systems.
Such a relation is first discussed theoretically some years ago in 
\cite{jumuro} and \cite{hemuro}, however without 
rigorous consideration of the biorthogonality of the eigenfunctions 
of $\ch$. The same holds true for the papers \cite{pastawski} on
the dynamical phase transition observed experimentally and
theoretically  in the spin swapping operation in atomic systems.
Only recently, the mixing of the wavefunctions at an EP 
and its relation to a dynamical phase transition is studied for an
open quantum system with more than two states  \cite{nearby2}. 
The states at both sides of the phase transition are not analytically
related to one another, meaning that this characteristic feature of 
any phase transition is fulfilled.
The theoretical results are compared to experimental results in the
recent review \cite{ropp}. 

In contrast to these differences between Hermitian and non-Hermitian 
quantum physics at and near to an EP, the Hamilton operator 
of the Schr\"odinger equation of an open quantum system is almost 
Hermitian, to a good approximation, far from an EP. Here the
eigenfunctions of the non-Hermitian Hamilton operator are almost 
orthogonal \cite{nearby1};  and the system can be described quite 
well by a standard Hermitian Hamilton operator. We underline however
that the Hamiltonian remains definitely non-Hermitian since the
non-Hermiticity arises solely from the fact that the function space of
the localized part of the system is a subsystem of the total function 
space\,: the localized part, we are interested in, is embedded into 
an extended environment of scattering wavefunctions, see
Fig. 1 in \cite{ropp}. The eigenfunctions of a 
non-Hermitian Hamilton operator are always biorthogonal
\cite{top} (see also Sect. 3.2 in \cite{ropp}). This includes  the 
case with well-separated resonance states as shown theoretically 
\cite{savin1} and experimentally \cite{savin2}.

It is the aim of the present paper to consider the influence of EPs
onto the  eigenfunctions (including their phases)
of a non-Hermitian Hamiltonian
in detail. We are interested, above all, in the behavior of the 
eigenfunctions and their phases in approaching an EP and, furthermore, 
in the role they play  in a dynamical phase transition. 

First, we sketch in Sect.
\ref{eigen} the characteristic features of the eigenvalues and
eigenfunctions of a non-Hermitian 2x2 Hamiltonian which are known 
in literature, and provide the definitions of typical values such as,
among others, the phase rigidity. This is a quantitative
expression for the biorthogonality of the wavefunction and thus,
ultimately, of the degree of opening of the system.
 
In the following Sect. \ref{eps}
we consider the  conditions for the appearance of EPs in different
systems  consisting of two states.
Analytical results are obtained and discussed for
cases with two EPs.  
Then, we consider  in Sect. \ref{adj} the more general case of a 
system with more than two states where a clustering of EPs 
may occur. In Sect. \ref{num} we provide numerical results for systems 
under typical  conditions and compare them with analytical 
results, if such results exist. We
discuss the obtained results in Sect. \ref{disc}. Some of the
results are expected and agree with our general understanding of open
quantum systems. Other results are completely unexpected. These
results are considered and discussed in detail. They allow us to
receive a deeper understanding of dynamical phase transitions. We
conclude the paper with some remarks on the stabilization of open
quantum systems due to the existence of EPs; and discuss the
possibility to describe them by means of a Hermitian Hamiltonian.

\section{Eigenvalues and eigenfunctions of the non-Hermitian Hamiltonian $\ch^{(2)}$}
\label{eigen}

In order to study the interaction of two states via the 
common environment it is most convenient to start from the symmetric
$2\times 2$ non-Hermitian matrix \cite{nearby1}
\begin{eqnarray}
{\cal H}^{(2)} = 
\left( \begin{array}{cc}
\varepsilon_{1} \equiv e_1 + \frac{i}{2} \gamma_1  & ~~~~\omega_{12}   \\
\omega_{21} & ~~~~\varepsilon_{2} \equiv e_2 + \frac{i}{2} \gamma_2   \\
\end{array} \right) 
\label{form1}
\end{eqnarray}
with $\gamma_i \le 0$ for decaying states
\cite{comment2}. The 
$\omega_{12}=\omega_{21}\equiv \omega$ stand for the coupling
matrix elements of the two states via the common environment which
are, generally, complex \cite{top}. The diagonal elements 
$\varepsilon_i$ of (\ref{form1}) contain the energies
$e_i$ and decay widths  $\gamma_i$ of the two states 
when $\omega_{ij} =0$, i.e. they are the two complex eigenvalues 
$ \varepsilon_{i}~(i=1,2)$ of the non-Hermitian operator 
${\cal H}_0^{(2)}$ that describes the system without any coupling 
of its states via the environment. In the present  paper, we take
$\omega_{12}=\omega_{21}\equiv \omega$, 
and the self-energy of the states is assumed to be included into
$\varepsilon_{1}$ and $\varepsilon_{2}$.  
We underline here that the 
Hamiltonian $\ch^{(2)}$ is completely non-Hermitian (see
also \cite{nearby1}) in 
difference to the many non-Hermitian operators used in  the literature
for the description of open quantum systems. These operators consist
mostly of a Hermitian part to which a non-Hermitian part is added as a
perturbation.  

The model (\ref{form1}) seems to be very simple. This is however
not true from a mathematical point of view. The point is that
singularities are involved in the model (the so-called EPs) 
which are known in mathematics for many years, see 
[3]. They are considered in physics only recently. They cause 
counterintuitive results \cite{top} which seem to be, at first glance, 
wrong. We will discuss them in the following. 

In the case  $\omega =0$,  the real part $E_k$ of the eigenvalues of
$\ch^{(2)}$ does not differ from the original energies $e_k$. It follows,
under this condition,
$E_{1,2}= \frac{1}{2}~(e_1+e_2) ~\pm ~\frac{1}{2}~(e_1-e_2) ~= ~e_{1,2} $.
A corresponding relation holds for the $\Gamma_k$ relative to the 
original $\gamma_k$.

Most interesting properties of $\ch^{(2)}$ are
the crossing points of two eigenvalue trajectories. Since here the two states
coalesce at one point, the influence of all the other states of the
system on the interaction of these two states can be neglected.
Therefore,  (\ref{form1}) describes the characteristics of open quantum
systems that may be related to these points, in spite of its small rank.

The eigenvalues of $\ch^{(2)}$ are, generally, complex and may be expressed as
\begin{eqnarray}
\ce_{1,2} \equiv E_{1,2} + \frac{i}{2} \Gamma_{1,2} = 
\frac{\varepsilon_1 + \varepsilon_2}{2} \pm Z ~; \quad \quad
Z \equiv \frac{1}{2} \sqrt{(\varepsilon_1 - \varepsilon_2)^2 + 4 \omega^2}
\label{int6}
\end{eqnarray}
where  $E_i$ and $\Gamma_i$ stand for the energy and width,
respectively, of the eigenstate $i$. Also here $\Gamma_i \le 0$ for
decaying states \cite{comment2}.
The two states may repel each other in accordance with Re$(Z)$,
or they may undergo width bifurcation in accordance with Im$(Z)$.
When $Z=0$ the two states cross each other at a point that 
is called usually {\it exceptional point} (EP) \cite{kato}. The EP is a
singular point (branch point) in the complex plane where the $S$-matrix has 
a double pole \cite{top}. 

The eigenfunctions of any non-Hermitian operator $\ch$ 
must fulfill the conditions $\ch|\Phi_i\rangle =
  {\ce}_i|\Phi_i\rangle$ and $\langle \Psi_i|\ch = {\ce}_i \langle
  \Psi_i|$ where  $\ce_i$ is an eigenvalue of $\ch$ and the vectors 
$ |\Phi_i\rangle$ and $\langle \Psi_i|$ denote its right and left
eigenfunctions, respectively \cite{top}. When $\ch$ is a Hermitian 
operator, the  $\ce_i$ are real, and we arrive at the
well-known relation $\langle \Psi_i| =  \langle \Phi_i|$. In this case, 
the eigenfunctions can be normalized by using the expression 
$\langle \Phi_i|\Phi_j\rangle$. 
For the symmetric non-Hermitian Hamiltonian $\ch^{(2)}$, however, we
have  $\langle \Psi_i| =  \langle \Phi_i^*|$. This means, that the
eigenfunctions are biorthogonal and have to be normalized by means of 
$\langle \Phi_i^*|\Phi_j\rangle$. This is, generally, 
a {\it complex} value, in contrast to the real value 
$\langle \Phi_i|\Phi_j\rangle$ of the Hermitian case. To smoothly
describe the transition from a closed system with discrete states, to
a weakly open one with narrow resonance states, we normalize the 
$\Phi_i$  according to 
\begin{eqnarray}
\langle \Phi_i^*|\Phi_j\rangle = \delta_{ij} 
\label{int3}
\end{eqnarray}
(for details see Sects. 2.2 and 2.3 of \cite{top}). 
It follows  
\begin{eqnarray}
 \langle\Phi_i|\Phi_i\rangle & = & 
{\rm Re}~(\langle\Phi_i|\Phi_i\rangle) ~; \quad
A_i \equiv \langle\Phi_i|\Phi_i\rangle \ge 1
\label{int4} 
\end{eqnarray}
and 
\begin{eqnarray}
\langle\Phi_i|\Phi_{j\ne i}\rangle & = &
i ~{\rm Im}~(\langle\Phi_i|\Phi_{j \ne i}\rangle) =
-\langle\Phi_{j \ne i}|\Phi_i\rangle 
\nonumber  \\
&& |B_i^j|  \equiv 
|\langle \Phi_i | \Phi_{j \ne i}| ~\ge ~0  \; .
\label{int5}
\end{eqnarray}

At the EPs, not only the eigenvalues of two states coalesce but also 
the two corresponding 
eigenfunctions of the non-Hermitian Hamilton operator $\ch^{(2)}$  
are the same, up to  a phase, 
\begin{eqnarray}
\Phi_1^{\rm cr} \to ~\pm ~i~\Phi_2^{\rm cr} \; ;
\quad \qquad \Phi_2^{\rm cr} \to
~\mp ~i~\Phi_1^{\rm cr}   \; .
\label{eif5}
\end{eqnarray}  
These relations follow from analytical as well as 
from numerical and experimental
studies, see  Appendix of \cite{fdp1}, Sect. 2.5 of \cite{top}
and Figs. 4 and 5 in \cite{berggren}.
We underline here that the coalescence of the two eigenvalues
of a non-Hermitian operator at an EP should not be confused with the 
well-known fact that two eigenstates of a Hermitian operator may be 
degenerate. The difference consists in the
relation (\ref{eif5}) between the two corresponding eigenfunctions
which does, of course, not hold for degenerate states. Furthermore, 
the eigenfunctions of a non-Hermitian operator are biorthogonal while
those of degenerate states are orthogonal.

According to (\ref{eif5}), the wavefunction $\Phi_1$ of the state 
$1$ jumps, at the EP, to   ~$\pm\, i\, \Phi_2$ \cite{top,comment}. 
This mathematical behavior of the eigenfunctions $\Phi_i$ at the singular EPs 
causes the main differences between the physics of Hermitian and non-Hermitian 
quantum systems. At an EP, $A_i \to \infty , ~ |B_i^j| \to \infty$,
and the influence of the environment
onto the system is extremely large \cite{top}.

In (\ref{int3}),  the complex value 
$\langle \Phi_i^*|\Phi_j\rangle $ is normalized to the real value 
$\delta_{ij}$ with the consequence that the relative phase between the biorthogonal eigenfunctions of two
neighbored states changes  in such a
manner that always Im$\langle \Phi_i^*|\Phi_j\rangle =0 $. 
A quantitative measure of this change is the so-called {\it phase rigidity} 
\begin{eqnarray}
r_k ~\equiv ~\frac{\langle \Phi_k^* | \Phi_k \rangle}{\langle \Phi_k 
| \Phi_k \rangle} ~= ~A_k^{-1} 
\label{eif11}
\end{eqnarray}
of the state $k$ which is defined by the ratio between biorthogonality
and orthogonality of the  wavefunctions $\Phi_k$.
For Hermitian systems for which 
$\langle \Phi_k^*|\Phi_k\rangle = \langle\Phi_k|\Phi_k\rangle$ holds,  
the phase rigidity is equal to unity. This is
an expression of the fact that the eigenfunctions of Hermitian
operators are orthogonal. For weakly decaying
systems, where  one has well-separated resonance states,
the wavefunctions are almost orthogonal, i.e. the degree of
biorthogonality is small.  Under such 
conditions, Hermitian quantum physics represents a reasonable 
approximation to the description of the open quantum system. 
However,  the wavefunction $\Phi_1$ jumps at the EP to $\pm \, i
\,\Phi_2$ and vice versa; and the phase rigidity does not vary
continuously at the EP also in this case. 
 
The Schr\"odinger equation with the non-Hermitian Hamilton 
operator ${\cal H}^{(2)}$ is equivalent 
to a Schr\"odinger equation with ${\cal H}_0^{(2)}$ and source term \cite{ro01}
\begin{eqnarray}
\label{form1a}
({\cal H}_0^{(2)} - \varepsilon_i) ~| \Phi_i \rangle  = -
\left(
\begin{array}{cc}
0 & \omega_{ij} \\
\omega_{ji} & 0
\end{array} \right) |\Phi_j \rangle \equiv W  |\Phi_j \rangle\; . 
\end{eqnarray}
This equation relates $\Phi_i$ to $\Phi_{j\ne i}$ in a non-trivial
manner  due to the source term. That means, two states $i$ and $j\ne i$
are coupled  via the 
common environment of scattering wavefunctions into which the system 
is embedded. It is  $\omega_{ij}=\omega_{ji}\equiv\omega$, and the
coupling between the states $i$ and $j \ne i$ vanishes when $\omega
\to 0$. 
The Schr\"odinger equation (\ref{form1a}) with source term can be
rewritten in the following manner \cite{ro01},
\begin{eqnarray}
\label{form2a}
({\cal H}_0^{(2)}  - \varepsilon_i) ~| \Phi_i \rangle  = 
\sum_{k=1,2} \langle
\Phi_k|W|\Phi_i\rangle \sum_{m=1,2} \langle \Phi_k |\Phi_m\rangle 
|\Phi_m\rangle \; . 
\end{eqnarray}
According to the biorthogonality  relations
(\ref{int4}) and (\ref{int5}) of the eigenfunctions of ${\cal H}^{(2)}$,  
(\ref{form2a}) is a nonlinear equation,   since 
$\langle \Phi_k |\Phi_m\rangle \ne 1$ for $k= m$ and 
$\langle \Phi_k |\Phi_m\rangle \ne 0$ for $k\ne m$. 
Most important part of the nonlinear contributions is contained in 
\begin{eqnarray}
\label{form3a}
({\cal H}_0^{(2)}  - \varepsilon_n) ~| \Phi_n \rangle =
\langle \Phi_n|W|\Phi_n\rangle ~|\Phi_n|^2 ~|\Phi_n\rangle \; .  
\end{eqnarray}
The nonlinear source term vanishes far from an EP where
$\langle \Phi_k|\Phi_{k }\rangle$ approaches $1$ and
$\langle \Phi_k|\Phi_{l\ne k }\rangle = - 
\langle \Phi_{l \ne k  }|\Phi_{k}\rangle$ approaches zero.
This follows from the normalization (\ref{int3})
which differs only a little from the standard normalization
$\langle \Phi_k|\Phi_{k }\rangle =  1 $ and
$\langle \Phi_k|\Phi_{l\ne k }\rangle = 0$ for a Hermitian 
Hamilton operator. Thus, the Schr\"odinger equation with source term 
is (almost) linear far from an EP, as usually assumed. It is however 
nonlinear in the neighborhood of an EP.

The nonlinear terms  in  (\ref{form3a}) 
cause, among others,  a mixing of the wavefunctions 
which can be expressed by
\begin{eqnarray}
\Phi_k=\sum_{l=1}^N b_{kl} \Phi_l^0 \; .
\label{int20}
\end{eqnarray}
where the  eigenfunctions 
$\Phi_k$ of ${\cal H}^{(2)}$ are represented in the
set of basic wavefunctions $\Phi_l^0$ of the operator ${\cal H}_0^{(2)}$
the non-diagonal matrix elements of which vanish.
For some numerical results see \cite{nearby1}. 
This mixing of the wavefunctions appears {\it additionally}
to other possible sources of mixing caused for some other reasons.

The eigenfunctions $\Phi_i$ and the eigenvalues $\ce_i$  
of $\ch^{(2)}$ contain global features that are 
caused by many-body forces  induced by the coupling
$\omega_{ik}$ of the states $i$ and $k\ne i$ via the environment. The
environment is the continuum of scattering wavefunctions and
has an infinite number of degrees of freedom.

\section{Exceptional points}
\label{eps}

We consider now the behavior that arises when the 
parametrical detuning of the two eigenstates of $\ch^{(2)}$ is varied, 
bringing them towards coalescence. 
According to (\ref{int6}), the  condition for coalescence reads 
\begin{eqnarray}
Z = \frac{1}{2} \sqrt{(e_1-e_2)^2 - \frac{1}{4} (\gamma_1-\gamma_2)^2 
+i(e_1-e_2)(\gamma_1-\gamma_2) + 4\omega^2} ~=~ 0 \; .
\label{int6i}
\end{eqnarray}
It follows that two interacting discrete states (with
$\gamma_1 = \gamma_2 =   0$ and $e_1 \ne e_2$) avoid always crossing since 
$\omega \equiv \omega_0$ and 
$\varepsilon_1 - \varepsilon_2$ are real in this case and the 
condition $Z=0$ can not be fulfilled,
\begin{eqnarray}
(e_1 - e_2)^2 +4\, \omega_0^2 &>& 0 \; .  
\label{int6a}
\end{eqnarray}
In this case, the EP can be found only by
analytical continuation into the continuum. This situation is called usually
avoided crossing of discrete states.
It holds also for narrow resonance states if $Z=0$ cannot be
fulfilled due to the small widths of the two states. 
The physical meaning of this result is very well known since many
years: the avoided crossing of two discrete states at a certain critical
parameter value \cite{landau} means that the
two states are exchanged at this  point, including their 
populations ({\it  population transfer}).  
 
When $\gamma_1 = \gamma_2$, and  $\omega = i\,\omega_0$ is 
{\it imaginary}, it follows from (\ref{int6i})  
\begin{eqnarray}
(e_1 - e_2)^2 -4\, \omega_0^2 &= &0 
~~\rightarrow ~~e_1 - e_2 =\pm \, 2\, \omega_0 
\label{int6b}
\end{eqnarray}
such that two EPs appear. It furthermore holds that
\begin{eqnarray}
\label{int6c}
(e_1 - e_2)^2 >4\, \omega_0^2 &\rightarrow& ~Z ~\in ~\Re \\
\label{int6d}
(e_1 - e_2)^2 <4\, \omega_0^2 &\rightarrow&  ~Z ~\in ~\Im 
\end{eqnarray}
independent of the parameter dependence of $e_{1,2}$.

Also in the case when the widths $\gamma_{1,2}$ are parameter
dependent, $e_1=e_2$ and $\omega$ is {\it real}, we have two EPs. 
Instead of (\ref{int6b}) to (\ref{int6d}) we have
\begin{eqnarray}
(\gamma_1 - \gamma_2)^2 -16\, \omega^2 &= &0 
~~\rightarrow ~~\gamma_1 - \gamma_2 =\pm \, 4\, \omega 
\label{int6e}
\end{eqnarray}
and
\begin{eqnarray}
\label{int6f}
(\gamma_1 - \gamma_2)^2 >16\, \omega^2 &\rightarrow& ~Z ~\in ~\Im \\
\label{int6g}
(\gamma_1 - \gamma_2)^2 <16\, \omega^2 &\rightarrow&  ~Z ~\in ~\Re \; .
\end{eqnarray}

Eqs. (\ref{int6c}) and (\ref{int6f}), respectively, 
describe the behavior away from the EPs, where the 
eigenvalues ${\cal E}_{k}$ only differ from the original ones 
through a contribution to the energy and width, respectively. 
The widths (or energies), in contrast, remain
unchanged, and this situation therefore corresponds to that of level
repulsion (or width bifurcation).
Eqs. (\ref{int6d}) and  (\ref{int6g}), in contrast, are relevant 
over the range between the two EPs and indicate that 
the resonance states  undergo width bifurcation (or level repulsion)
according to Im$(Z)\ne 0$ (and Re$(Z)\ne 0$, respectively).
The bifurcation (or level repulsion) starts in the neighborhood of
either one of the
EPs, and grows to reach a maximum value at the midpoint between
them (even though $\omega_0$ and $\omega$, respectively, remain fixed). 
The condition for maximum width bifurcation (or level repulsion) is
fulfilled at the crossing point $e_1 = e_2$ (and $\gamma_1 =
\gamma_2$, respectively).
Physically, the bifurcation implies that {\it different time scales}
may appear in the system, while the states are nearby to one another
in energy (for details see \cite{fdp1}). In an analogous manner,
level repulsion of states with 
similar lifetimes causes a {\it separation of the states in energy}.   
For an illustration of these analytical results see the numerical 
results presented in Sect. \ref{num2}, Fig. \ref{fig1} and Fig. 
\ref{fig3} left panel.

Under more realistic conditions,  $\omega$ is complex, and simple 
analytical results like (\ref{int6b}) to (\ref{int6g}) cannot be 
obtained. For this case, we 
will provide some results of numerical studies in 
Sect. \ref{num2}. In order to understand the meaning of these
numerical results, the analytical relations (\ref{int6b}) to
(\ref{int6g}) and their representation in Figs. \ref{fig1} and Fig. 
\ref{fig3} left panel are very helpful.

In any case, the parametric
dependence of the eigenvalues $\ce_{k}$ is non-analytical in the
vicinity of an EP, with the widths $\Gamma_k$, in particular, showing
variations that are  inconsistent with the predictions of Fermi's
golden rule (according to which the widths should increase with
increasing coupling strength of the system to the environment;
for details see \cite{top}). At these points, the 
influence of the environment onto the system properties
is extremely strong. In our case, the environment is the continuum 
of scattering wavefunctions which gives  
to the eigenstates of $\ch^{(2)}$ a finite lifetime.

It follows from the normalization
condition  (\ref{int3}) that 
$\langle \Phi_k | \Phi_k \rangle  \to \infty$ such that  
$r_k \to 0$ \cite{top}
when an EP is approached. In other words, the
relative phase of the two eigenfunctions changes dramatically 
when the crossing point is approached.
Most significantly, as understood from analytical studies, 
as well as from numerics and experiment (see \cite{top,fdp1,comment}) is, 
that, in the  vicinity of the EP,
the eigenfunctions differ from one another by only a phase,
see (\ref{eif5}). 
In a recent theoretical study on a microwave cavity \cite{berggren}, 
the relations  (\ref{eif5}) could be confirmed by the observation that
the real and imaginary components of two nearby eigenstates are 
``swapped'', under the influence of an EP, in complete agreement with 
(\ref{eif5}). The non-rigidity of the phases follows,
of course, directly from the fact that $\langle\Phi_k^*|\Phi_k\rangle$
is a complex number (in difference to the norm
$\langle\Phi_k|\Phi_k\rangle$, which is a real number)
so that the normalization condition (\ref{int3}) can be fulfilled only
by the additional  requirement 
Im$\langle\Phi_k^*|\Phi_k\rangle =0$ (corresponding to a rotation away
from the complex plane). Here, the two different states of the system  
develop, according to (\ref{eif5}), a coupling through the continuum,
a quantitative measure of which is the phase rigidity.
Thus, the biorthogonality of the eigenfunctions
$\Phi_k$ causes perceptible physical effects in the neighborhood of an EP.

Generally speaking, the phase rigidity takes values between zero and
one, with the value $r_k=1$  for Hermitian systems.
Near to an EP in a non-Hermitian system, however, 
the two eigenfunctions differ from one another only by a phase,  
according to  (\ref{eif5}), so that
$r_k \ll 1$. This non-rigidity of the eigenfunction phases is the most
important difference between Hermitian and non-Hermitian
eigenfunctions. Its meaning cannot be overestimated.
On the one hand, the lack of phase rigidity
near to an EP leads very naturally to the appearance of nonlinear
effects in the Schr\"odinger equation (\ref{form3a}) with source term 
which describes an open quantum system. On the other hand,
the impact of the environment  on the (localized) system 
is extremely strong at the EP. Since the environment is the continuum  
of scattering wavefunctions with an infinite number of 
degrees of freedom,  this impact may induce phase transitions
as discussed in \cite{nearby2}.

\section{Clustering of exceptional points}
\label{adj}

According to mathematical studies, more than two eigenvalues of a
non-Hermitian operator $\ch$ may cross in one point, the so-called  
higher-order EP. This crossing point is however a point in the 
continuum and therefore of measure zero \cite{nearby2}. 
In this respect, it does not differ from the second-order EP which 
is the crossing point of two eigenvalues 
considered in the foregoing Sect. \ref{eigen}. 
That means, a higher-order EP can {\it not directly} be identified 
in a realistic physical system. 
Nevertheless, it influences the dynamics of an  open quantum system 
in a similar manner as 
a second-order EP does; see the discussion  in the foregoing 
Sect. \ref{eigen} for a two-level system.

In \cite{nearby2}, the influence of a {\it third} state
onto the two eigenvalues and eigenfunctions of a non-Hermitian Hamilton 
operator that cross at an EP, is investigated. As a result, more than 
two states of a realistic physical system are unable to coalesce at 
one point since, in a certain finite parameter range around the
original second-order EP, the wavefunctions of the two states are
mixed. When the third state approaches this parameter range,
it crosses or avoids crossing therefore with
states that differ from the original two states. 
Accordingly, new EPs appear and the areas of influence of different 
EPs overlap. Altogether, the different EPs 
amplify, collectively, their impact onto physical values; and
the wavefunctions of all states
are strongly mixed in the basic wavefunctions 
$\Phi_i^0$ of ${\cal H}_0$.
This effect is nothing but some {\it clustering of EPs}, wherewith the 
characteristic fact is expressed that
the ranges of the influence of different second-order EPs 
overlap in a finite parameter range around a higher-order EP.

In the following, we will study the mixing of the wavefunctions and, 
above all, the phase rigidity defined in, respectively, (\ref{int20}) 
and  (\ref{eif11}), in the case of clustering of EPs. To this aim we 
consider the non-Hermitian  Hamiltonian 
\begin{eqnarray}
\label{f1}
\ch^{(n)}=\left(
\begin{array}{cccc}
\epsilon _{1}=e_{1}+i\frac{\gamma _{1}}{2} & \omega _{12} &\ldots &\omega _{1n} \\
\omega _{21} & ~~\epsilon _{2}=e_{2}+i\frac{\gamma _{2}}{2}~~ & 0 & 0 \\
\ldots & 0 &\ldots & 0 \\
\omega _{n1} & 0 & 0 & ~~\epsilon _{n}=e_{n}+i\frac{\gamma _{n}}{2}~~
\end{array}
\right) 
\end{eqnarray}
with $n=3$ or 4  nearby states coupled to one common continuum (the
first channel). As in (\ref{form1}), the $e_i$ and $\gamma_i$ denote
the energies and widths, respectively, of the $n$ states without
account of the interaction of the different states via the environment.   
The $\omega_{ij} = \omega_{ji}$ simulate the interaction of the two
states $i$ and $j$ via the common environment. 
In the simulation (\ref{f1}), we
used the doorway concept used in nuclear physics\,: the $n$ states
with the decay widths $\gamma_{i}/2$ can be simulated by one doorway
state with large decay width $\gamma_1$ and $n-1$ states with small 
(almost vanishing) decay widths $\gamma_{i\ne 1}$. Then (according to
the doorway concept), the doorway
state is coupled to both the environment and the remaining $n-1$ states, 
while the remaining states are coupled to the environment only via 
the doorway state (due to their small decay widths and the fact that
they are distant from EPs). The coupling strength $\omega$
between system and environment is not varied in our calculations, and the
number of parameters for the widths and energies of all $n$ states 
is $2n$ \cite{mahaux}. 

The eigenvalues of (\ref{f1}) can be obtained in analogy to 
(\ref{int6}). 
The eigenfunctions are biorthogonal. We normalize them according 
to (\ref{int3}). Further  (\ref{eif5}) holds at an EP. The values $A_i$
and $|B_i^j|$ defined in (\ref{int4}) and (\ref{int5}), respectively,
express how near the system is to an EP at the considered parameter
value. In the numerical calculations, these values can be seen
directly by studying the mixing coefficients $|b_{ij}|$ defined in
(\ref{int20}).  Also the corresponding phase rigidities $r_i$ of 
the different states can be determined numerically by using 
(\ref{eif11}). 

Near to the different EPs, the Schr\"odinger equation contains 
nonlinear contributions according to (\ref{form3a}) due to the
source term that describes the coupling between system and environment.         
Accordingly, the whole parameter range in which a clustering of EPs 
occurs,  is controlled by
nonlinear contributions to the Schr\"odinger equation, the values of
which vary because of their dependence on the concrete parameter value.
They vanish only far from this regime with a clustering of  EPs.

\section{Numerical results} 
\label{num}

\subsection{$N=2$ states}
\label{num2}

\begin{figure}[ht]
\begin{center}
\includegraphics[width=6cm,height=14cm]{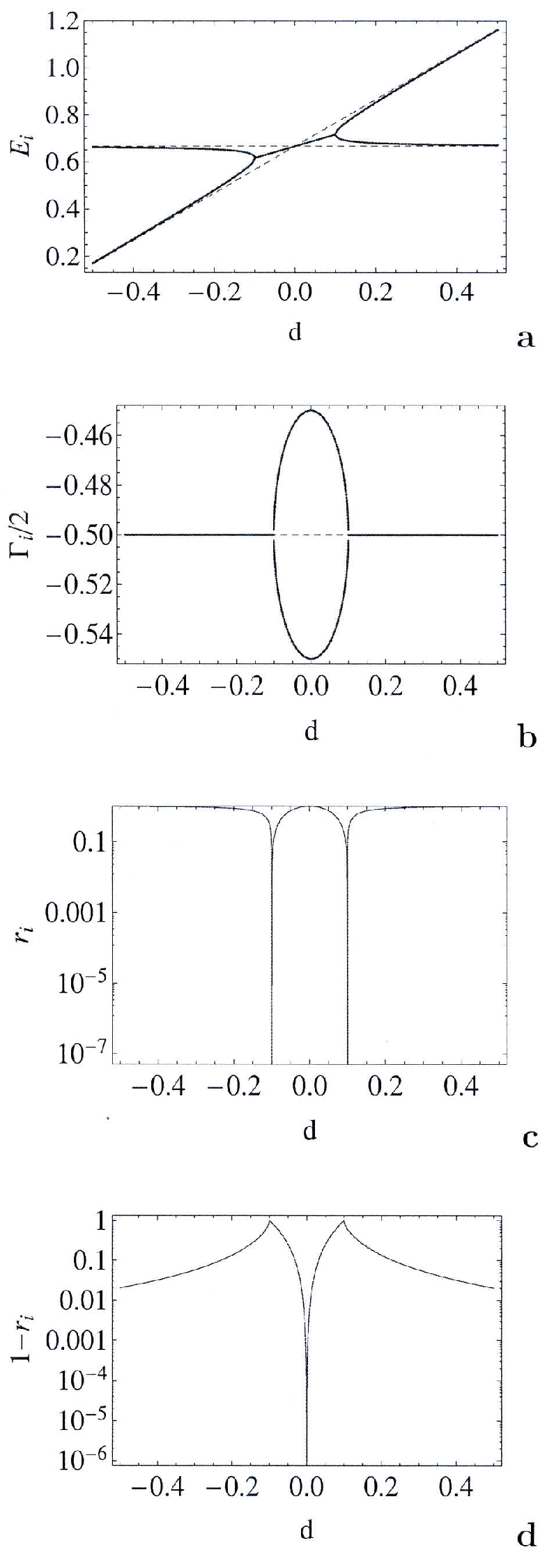}
~~~~\includegraphics[width=6cm,height=14cm]{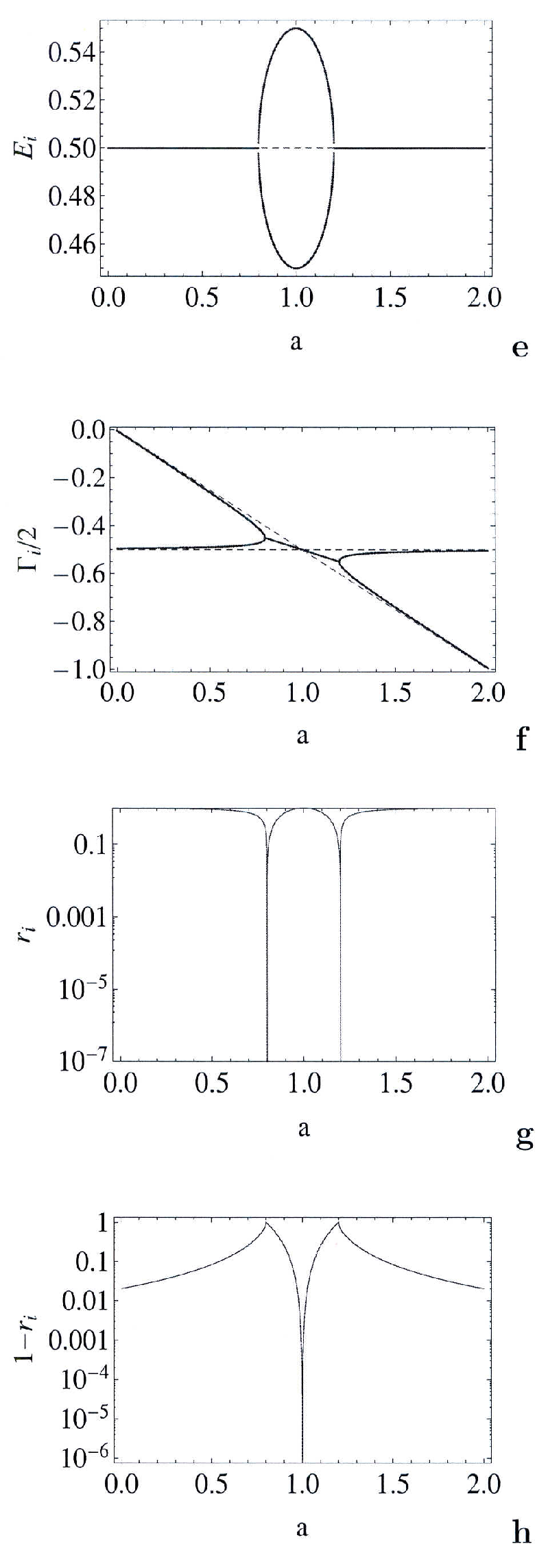}
\end{center}
\caption{\footnotesize
Energies $E_i$ (full lines) (a,e), widths $\Gamma_i/2$ (b,f), phase
rigidity $r_1 = r_2$   (c,g), and $1-r_k$  (d,h)   
of the two eigenfunctions of $\ch^{(2)}$ as a function of the
distance  $d$
between the two unperturbed energies $e_1$ and $e_2$. The parameters
are given, respectively, in the first and second row of 
Tab. \ref{tab1}. The dashed lines 
in (a,e) and (b,f) show, respectively, the $e_i$ and $\gamma_i/2$ 
trajectories.
}
\label{fig1}
\end{figure}

\begin{figure}[ht]
\begin{center}
\includegraphics[width=6cm,height=14cm]{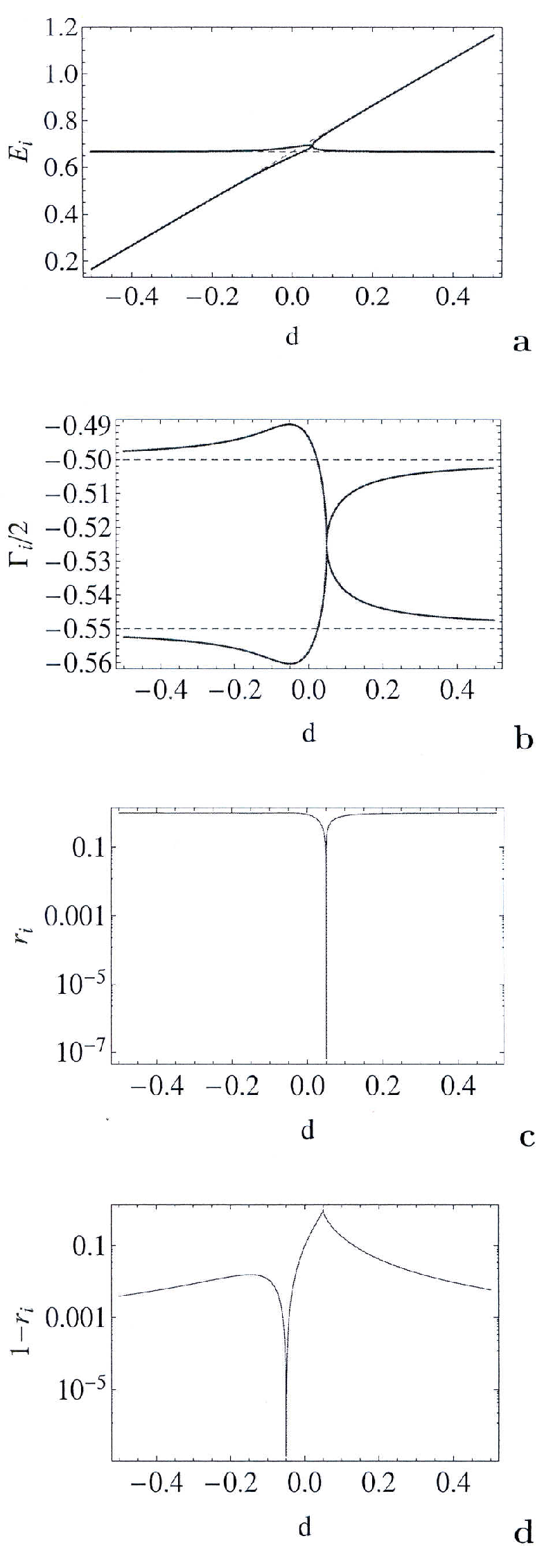}
~~~~\includegraphics[width=6cm,height=14cm]{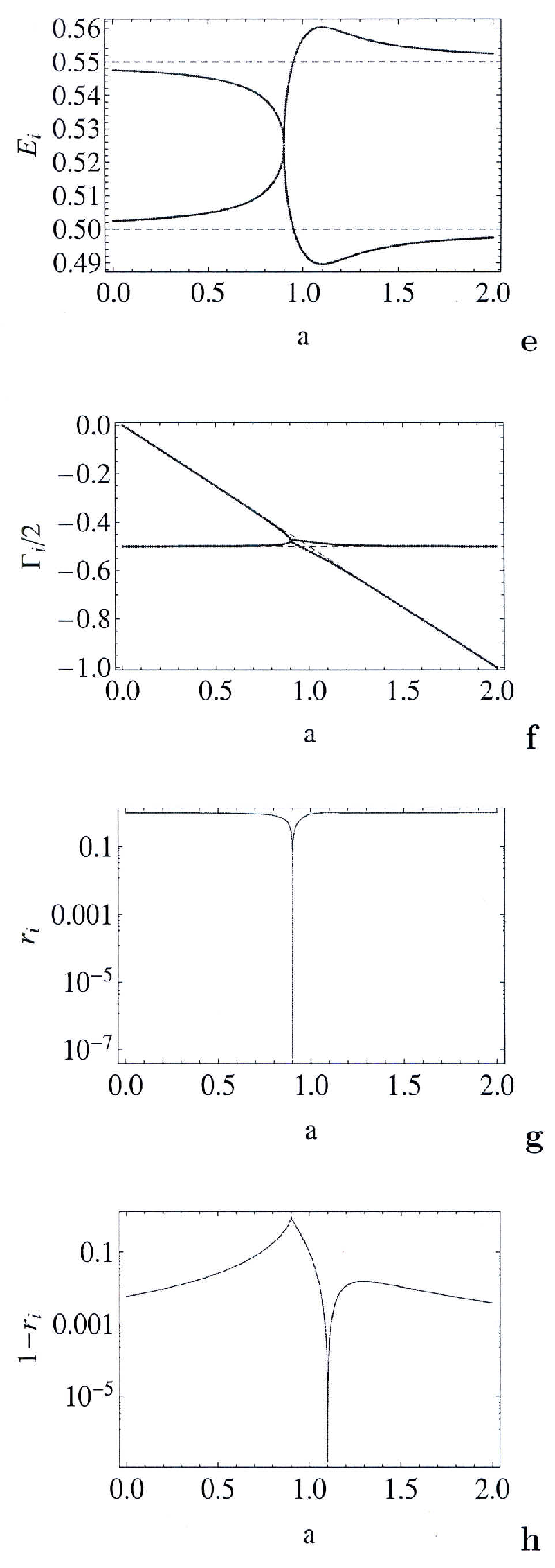}
\end{center}
\caption{\footnotesize
The same as Fig. \ref{fig1} but with the parameters
given, respectively, in the third and fourth row of Tab. \ref{tab1}.
}
\label{fig2}
\end{figure}

\begin{figure}[ht]
\begin{center}
\includegraphics[width=6cm,height=14cm]{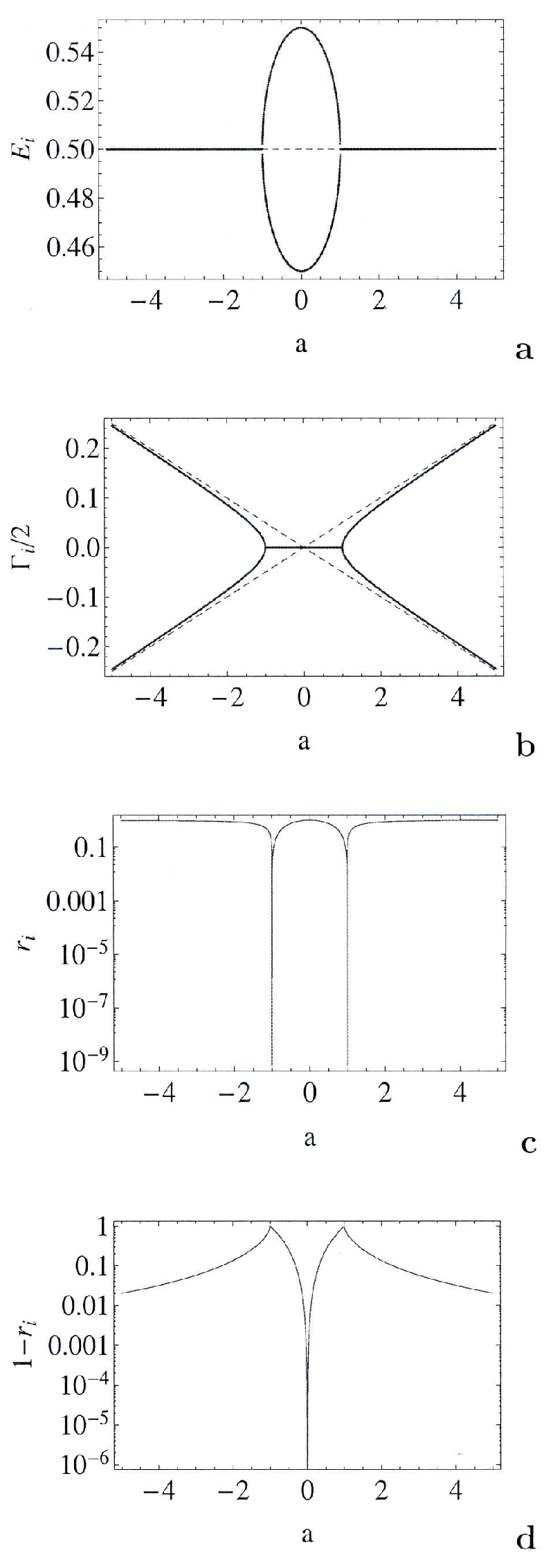}
~~~~\includegraphics[width=6cm,height=14cm]{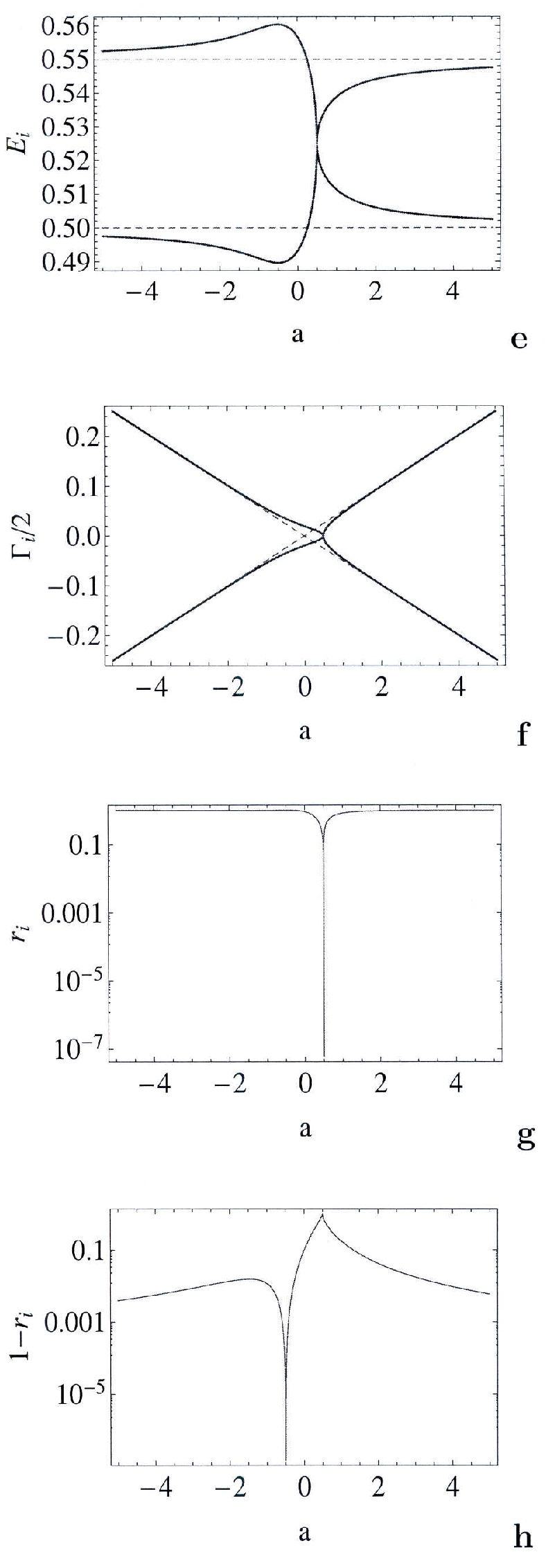}
\end{center}
\caption{\footnotesize
The same as Fig. \ref{fig1} but with the parameters
given, respectively, in the fifth and sixth row of Tab. \ref{tab1}.
}
\label{fig3}
\end{figure}

\begin{figure}[ht]
\begin{center}
\includegraphics[width=13cm,height=17.5cm]{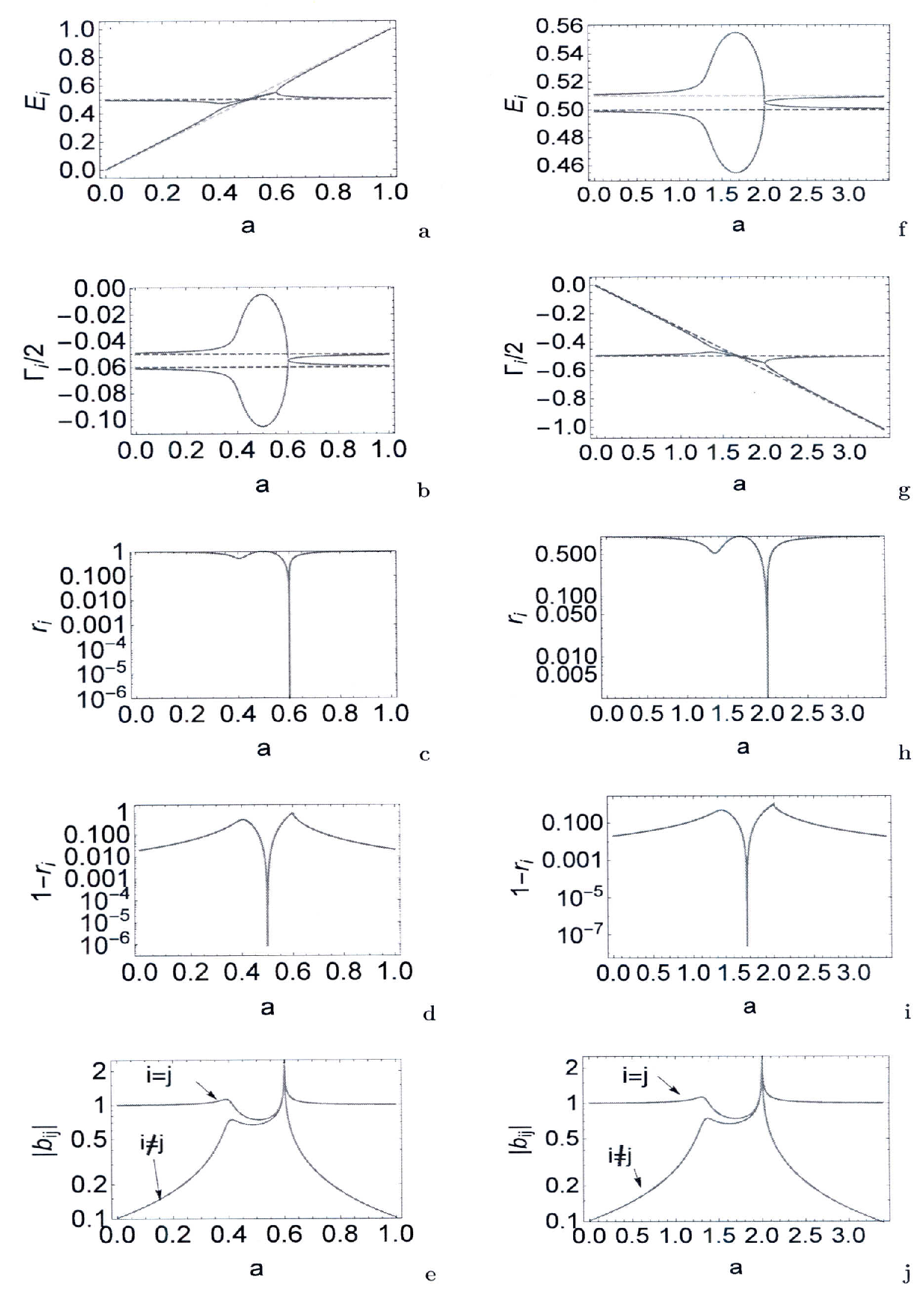}
\end{center}
\caption{\footnotesize
The same as Fig. \ref{fig1} but with the parameters
given, respectively, in the seventh and eighth row of Tab. \ref{tab1}.
In (e) and (j), the mixing coefficients $|b_{ij}|$ of the wavefunctions, 
defined by (\ref{int20}), are shown.}
\label{fig4}
\end{figure}

In Figs. \ref{fig1} to \ref{fig4}, we show the results of numerical
calculations performed with the parameters given in Table \ref{tab1}.
Most impressive is that {\it all results for the phase rigidity show the
same behavior} in spite of the different parameters and the fundamental
differences in the eigenvalue pictures. In {\it all} cases, the phase
rigidity $r_k$ approaches the value $r_k \to 0$ at the
position of the EP while it approaches sharply the value $r_k \to 1$ when  
width bifurcation and level repulsion, respectively, is maximum.   
These changes occur {\it without} any changes of the coupling strength
$\omega$ between system and environment as can be seen from 
the parameter values given Tab. \ref{tab1}.

\begin{table}
\caption{The parameters used in the calculations}
\vspace{.2cm}
\begin{tabular}{||c|c|c|c|c|c||}
\hline
\hline
~~~~~~Figure~~~~~~ & $~~~~~~e_1~~~~~~$ & $~~~~~~e_2~~~~~~$ & $~~~~~\gamma_1 /2~~~~~$ & $~~~~~\gamma_2 /2~~~~~$ & $~~~~~~~~~~\omega~~~~~~~~~~$  \\
\hline
\hline
Fig. \ref{fig1}.a--d &$2/3$ & $2/3+d$ & $-0.5$ & $-0.5$ & 0.05~i \\
\hline
Fig. \ref{fig1}.e--h& $1/2$ & $1/2$ & $-0.5$ & $-0.5~a$ & 0.05  \\
\hline
\hline
Fig. \ref{fig2}.a--d & $2/3$ & $2/3+d$ & $-0.5$ & $-0.55$ & 0.025\,(1+i) \\
\hline
Fig. \ref{fig2}.e--h & $0.55$ & $0.5$ & $-0.5$ & $-0.5~a$ & 0.025\,(1+i)  \\
\hline 
\hline
Fig. \ref{fig3}.a--d & $0.5$ & $0.5$ & $0.05~a$ & $-0.05~a$ & 0.05  \\
\hline
Fig. \ref{fig3}.e--h & $0.55$ & $0.5$ & $0.05~a$ & $-0.05~a$ & 0.025\,(1+i)  \\
\hline
\hline
Fig. \ref{fig4}.a--e & $0.5$ & $a$ & $-0.05$ & $-0.06$ & $0.05\,(\frac{1}{10}+i)$  \\
\hline
Fig. \ref{fig4}.f--j & $0.5$ & $0.51$ & $-0.5$ & $-0.3~a$ & $0.05\,(1+\frac{1}{10}\,i)$  \\
\hline
\hline
\end{tabular}
\label{tab1}
\end{table}

Results for examples with two EPs  
according to Eqs. (\ref{int6b}) to (\ref{int6d}) and     
(\ref{int6e}) to (\ref{int6g}), respectively, are shown 
in Fig. \ref{fig1}. The numerical results
agree with the analytical ones\,: the phase rigidity
approaches zero at the two EPs; while in between the EPs, we see 
width bifurcation in the first case (Fig. \ref{fig1}.b) and level
repulsion in the second case (Fig. \ref{fig1}.e). The result $r_k \to
0$ at every EP corresponds to the expectation of theory \cite{top}. 
However, there is  an unexpected sharp transition to $r_k \to 1$
at the point of maximum width bifurcation or maximum level repulsion.
This means that here the
two eigenfunctions of $\ch^{(2)}$ are (almost) orthogonal to one
another. Far from the critical region, the phase rigidity approaches
the value $1$ according to the fact that the influence of the
environment onto the system can be neglected, to a good approximation,
far from EPs.

The results for the more realistic case with complex coupling strength
$\omega$ are shown in Fig. \ref{fig2}. Here, only one EP appears. In a 
finite parameter distance from the EP, we see maximum width bifurcation 
(Fig. \ref{fig2}.b) and
maximum level repulsion (Fig. \ref{fig2}.e), respectively. Again,
$r_k\to 0$ at the EP and $r_k \to 1$ at maximum width bifurcation or
maximum level repulsion.  

The results in Fig. \ref{fig3}  show the eigenvalue
and phase rigidity pictures for the case when not only loss (as in
Figs. \ref{fig1} and \ref{fig2}) appears but also gain is a possible
process. Fig. \ref{fig3} left panel shows the case with balanced
loss and gain, corresponding to $\Gamma_k = 0$ in the 
finite parameter range  between the two EPs (see Fig. \ref{fig3}.b). 
Formally, this case is  similar to those  discussed 
recently in many papers related to non-Hermitian operators with $\cpt$ 
symmetry whose eigenvalues  are real 
in a finite parameter range, see e.g. \cite{bender,bender2}. The 
$\cpt$-symmetry breaking is caused by  EPs.  
 
The results shown in Fig. \ref{fig3} left panel have the same
characteristic features 
as those shown in Fig. \ref{fig1} right panel. The same holds true
when the coupling strength  $\omega$ is complex (Fig. \ref{fig3} right 
panel as compared to Fig.  \ref{fig2} right panel). 

In Fig. \ref{fig4}, some results are shown with, respectively,  
almost imaginary (left panel) and almost real (right panel) coupling 
strength $\omega$. We see again the characteristic sharp transition
$r_k \to 0$ in approaching  the EP and $r_k \to 1$  at another
parameter value at which we have maximum width bifurcation and
maximum level repulsion, respectively. Additionally, we show 
in Fig. \ref{fig4} the mixing of the
wavefunctions expressed by the coefficients $|b_{kl}|$ which are 
defined in 
(\ref{int20}). At the EP, $|b_{kl}| \to \infty$ as shown in
\cite{top}. As in the other figures with complex coupling strength
$\omega$ (Figs. \ref{fig2} and \ref{fig3} right panel),
there is only one EP. The point of maximum 
width bifurcation and maximum level repulsion, respectively, 
appears at a finite parameter distance from the EP. In this 
parameter region, the two wavefunctions are strongly mixed. 
The mixing remains when $r_k \to 1$ is approached. Beyond $r_k \to 1$,
we see the hint to a nearby EP ($0< r_k < 1$) which limits the 
extension of the total critical parameter region.
Beyond this critical parameter
region, the wavefunctions approach their original orthogonal
character. The physical meaning of the mixing of the wavefunctions 
under the influence of EPs is discussed in detail in \cite{nearby1,nearby2}.
Here, we underline only that the wavefunctions are mixed when 
$r_k \to 1$, i.e. when they are almost orthogonal in the critical
parameter region.

\subsection{$N>2$ states}
\label{num3}

\begin{figure}[ht]
\begin{center}
\includegraphics[width=13cm,height=15cm]{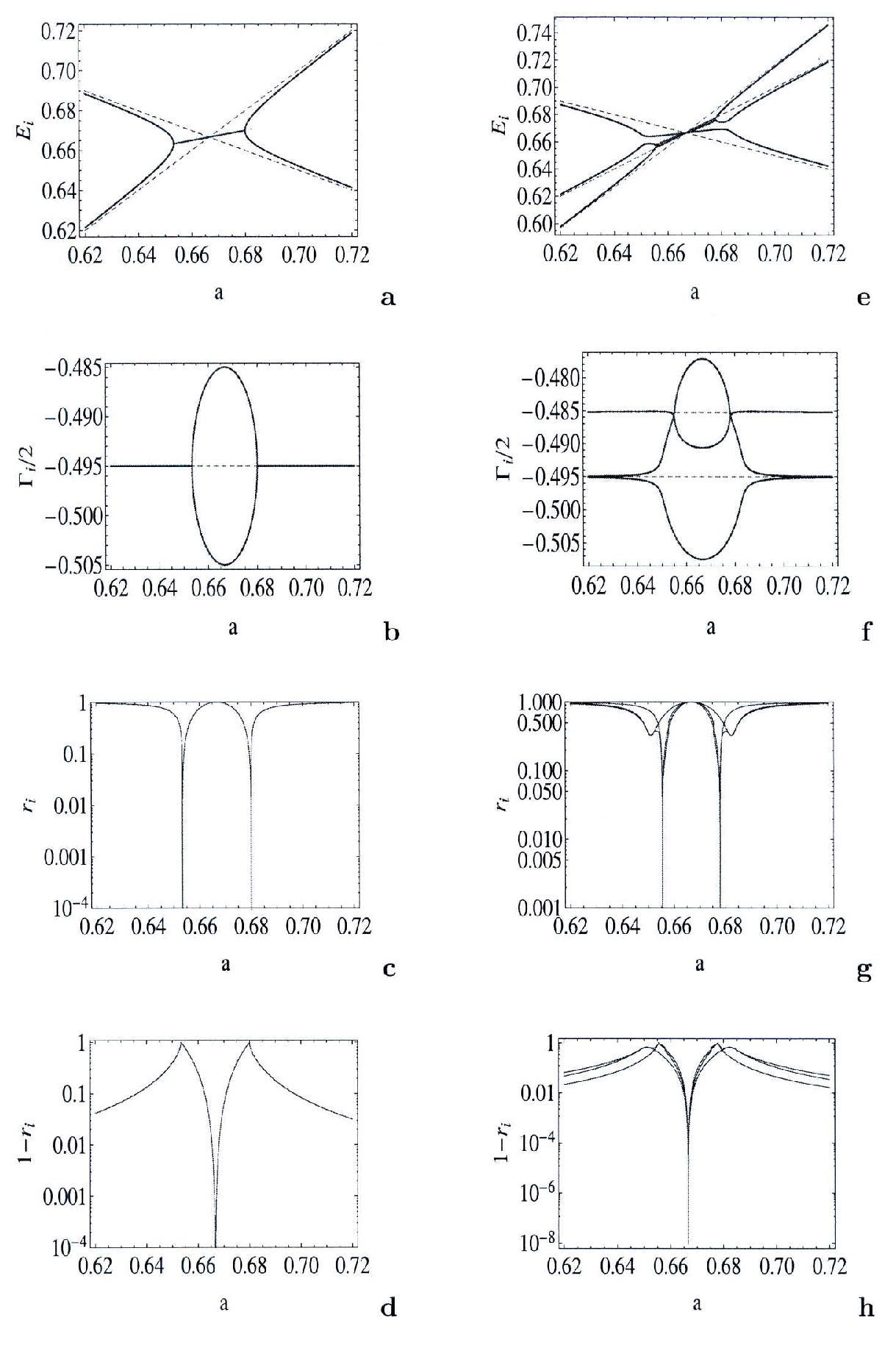}
\end{center}
\caption{\footnotesize
Energies $E_i$ (full lines) (a,e), widths $\Gamma_i/2$ (b,f), phase
rigidity $r_i$   (c,g), and $1-r_i$  (d,h)   
of the two eigenfunctions of $\ch^{(2)}$ 
(left panel) and of the three eigenfunctions of
$\ch^{(3)}$ (right panel), respectively,
as a function of the parameter $a$.
The dashed lines in (a,e) and (b,f) show, respectively, the $e_i$ 
and $\gamma_i/2$ trajectories. The parameters are 
$\omega = 0.01 \, i$; ~$e_1=1-1/2~a; ~~e_2=a$; ~~$e_3=-1/3+3/2~a$ (e-h);
 ~~$\gamma_1/2= \gamma_2/2 =- 0.495$; ~~$\gamma_3/2=- 0.4853$ (e-h).
}
\label{fig5}
\end{figure}

\begin{figure}[ht]
\begin{center}
\includegraphics[width=13cm,height=15cm]{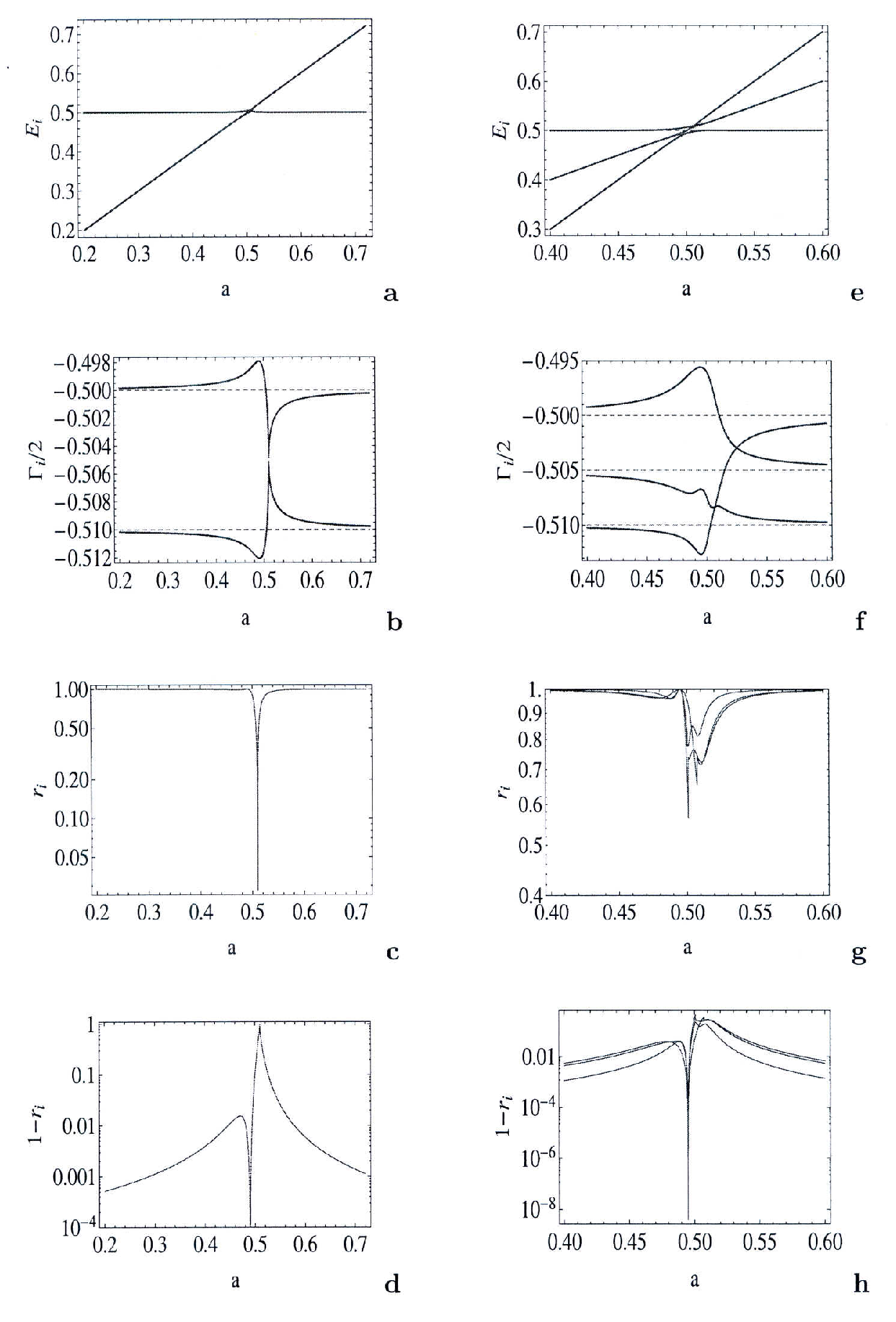}
\end{center}
\caption{\footnotesize
The same as Fig. \ref{fig5} but $\omega=0.005(1+i)$ and
$~~\gamma_{1}/2=-0.5;~~\gamma_{2}/2=-0.51;~~e_1=0.5;
~~e_2=a $ (left panel) and 
$~~\gamma_{1}/2=-0.5;~~\gamma_{2}/2=-0.505;~~\gamma_{3}/2=-0.51;
~~e_1=0.5;~~e_2=a;~~e_3=2a-0.5$ (right panel).
}
\label{fig6}
\end{figure}

\begin{figure}[ht]
\begin{center}
\includegraphics[width=12cm,height=14.5cm]{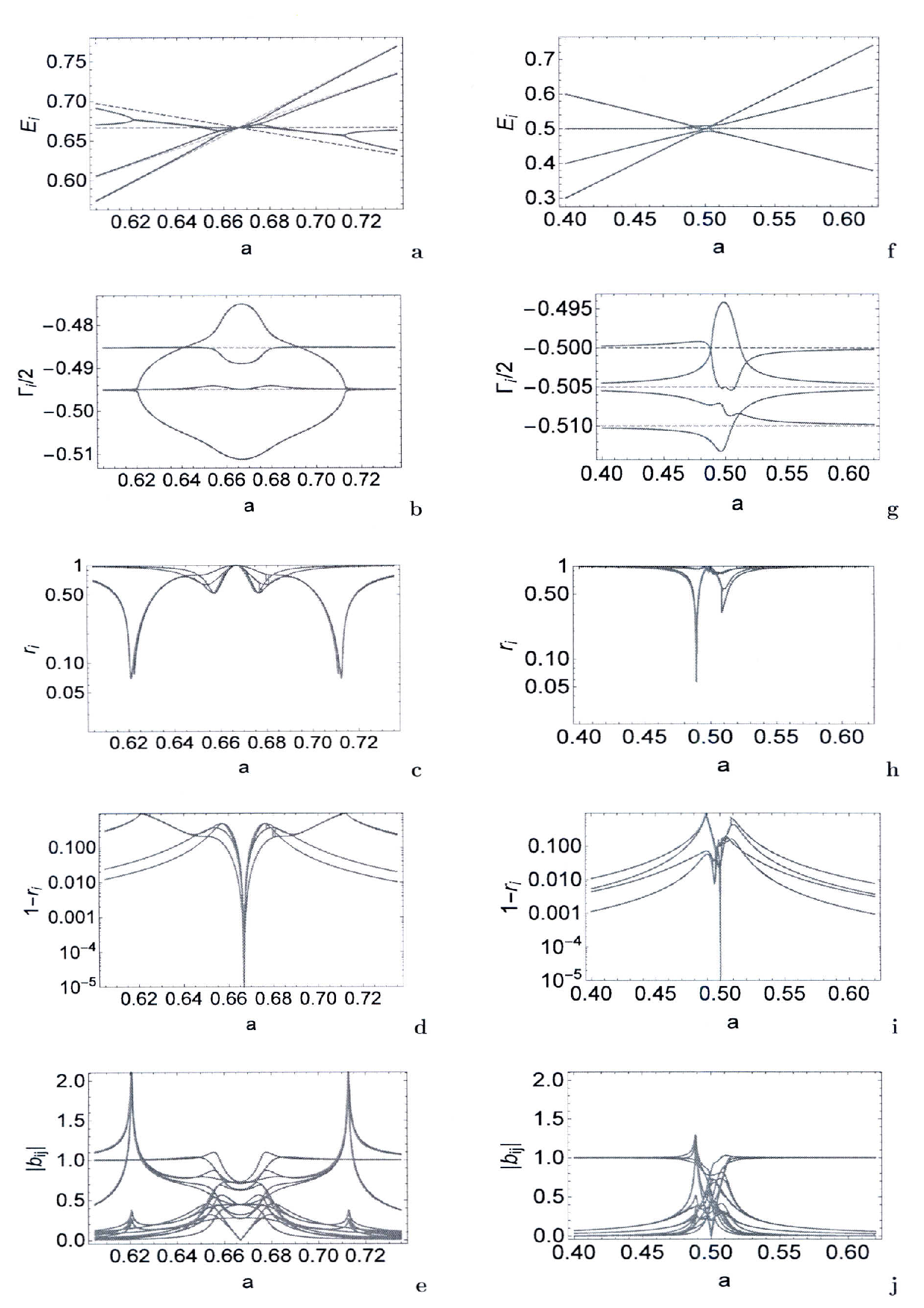}
\end{center}
\caption{\footnotesize
Energies $E_i$ (full lines) (a,f), widths $\Gamma_i/2$ (b,g), phase
rigidity $r_i$   (c,h),  $1-r_i$  (d,i), and 
mixing coefficients $|b_{ij}|$ of the wavefunctions, 
defined by (\ref{int20}), of the four eigenfunctions of $\ch^{(4)}$ 
as a function of the parameter $a$.
The dashed lines in (f,g) show, respectively, 
the $e_i$ and $\gamma_i/2$ trajectories.
The parameters are
$\omega = 0.01 \, i$;
~~$e_1=1-a/2; ~~e_2=a;~~e_3=-1/3+3/2 ~a; ~~e_4= 2/3;
~~ \gamma_1/2 = \gamma_2/2=-0.4950; ~~\gamma_3/2=-0.4853; 
~~ \gamma_4/2=-0.4950$ ~~(left panel) and
$\omega = 0.005 \, (1+i)$;
~~$e_1=0.5; ~~e_2=a;~~e_3=2a-0.5; ~~e_4= 1-a;
~~ \gamma_1/2 = -0.5; ~~\gamma_2/2=-0.505; ~~\gamma_3/2=-0.51; 
~~ \gamma_4/2=-0.505$ ~~(right panel).
}
\label{fig7}
\end{figure}

In Fig. \ref{fig5}, we show the influence of a "third" state onto the
eigenvalue picture and the phase rigidity around an EP in a system
that is  symmetric around the crossing point of the $e_i$
trajectories. New EPs can be identified
in the three-level case in the eigenvalue pictures and,
correspondingly, in the values of the phase rigidity ($r_i < 1$). The
two EPs near to the crossing point of the $e_i$ trajectories are well
expressed, and we see clear hints to  the existence of two other
distant EPs. Most interesting result is the
sharp transition from a reduced value  $r_i < 1$ of the phase rigidity
(which is characteristic of an EP in the neighborhood) back to  $r_i
\to 1 $ at maximum width bifurcation. This abrupt transition occurs
in both, the two-level system and the three-level one. At this
parameter value, the eigenfunctions of the non-Hermitian Hamiltonian
$\ch^{(2)}$ as well as those of $\ch^{(3)}$ become almost orthogonal.
At and around this critical parameter value, the wavefunctions of the
states are strongly mixed in both, the two-level case and the
three-level case \cite{nearby2}.

Fig. \ref{fig6}  shows results for a more realistic case with complex
coupling strength $\omega$ between system and environment. In the
three-level case, hints to the existence of different (distant) EPs
can be seen. The sharp  transition $r_i \to 1 $ at maximum width
bifurcation however appears very clearly not only in the two-level
case but also in the three-level case.

The results shown in the Figs. \ref{fig5} and \ref{fig6}
characterize the generic behavior of many-level open quantum systems
under the condition that the width bifurcation is maximum. The results
are confirmed by those which we received in many other calculations
performed with different parameters (including for systems which allow
loss and gain); or with a larger number $N$ of states.
For illustration we show numerical results obtained for $N=4$ states
in Fig. \ref{fig7}. Again we see hints to several EPs as well as the
sharp transition of $r_i$ to a value  almost 1 at a critical parameter
value. Here, the wavefunctions of all four states are strongly mixed.

\section{Discussion of the results} 
\label{disc}

\subsection{$N=2$ states}
\label{disc2}

Most surprising result of our study is the strong parameter dependence
of $r_k$ in a certain critical region around an EP. The variation $r_k \to 0$ 
in approaching an EP is expected from Eq. (\ref{eif5}).   The rapid
variation  $r_k \to 1$ when, respectively, the maximum width
bifurcation and level repulsion is approached, is however unexpected. The 
width bifurcation  or level repulsion  
starts at the EP without any enhancement of the coupling strength 
between system and environment. It is driven exclusively by the
nonlinear source term of the Schr\"odinger equation (\ref{form3a})
which  describes the open quantum system.
When  $r_k \to 1$ the wavefunctions of the two states, the 
eigenfunctions of which coalesce (up to a phase factor) at the EP 
(correspondingly to 
$r_k \to 0$), become almost orthogonal to one another. The
transition from    $r_k \to 0$ (at the EP) to  $r_k \to 1$ 
(where width bifurcation and level repulsion, respectively, 
is maximum) occurs 
suddenly as  function of the varied parameter 
in {\it all} our calculations.
The wavefunctions of the two states behave smoothly, i.e. they remain 
mixed also when $r_k \to 1$. 

The physical meaning of this result consists in the fact that
a stabilization of the localized part of the system occurs  when
the interaction $\omega$ between system and environment is strong
enough, i.e. when it is of the same order of magnitude as the 
widths $\gamma_i /2$. Then, 
in the case of Eqs. (\ref{int6b}) to (\ref{int6d}), one of the two
states receives a very short lifetime due to width bifurcation, and
becomes almost indistinguishable
from the  states of the environment. Although this process seems
to be reversible according to the figures for the eigenvalues, 
this is in reality not the
case. The processes occurring in approaching $r_k\to 1$, take place,
as mentioned above, by means of the nonlinear source term of 
the Schr\"odinger equation near an EP.  Due to these 
processes, the long-lived state has ``lost'' its short-lived   
partner with the consequence that the two original states cannot be 
reproduced. This evolution is therefore irreversible. The 
long-lived state is more stable than the original one, and the system
as a whole (which has lost one state)  is more stable than originally. 
The wavefunction of this long-lived state is mixed in those of the
original states.

The stabilization of the system due to  Eqs. (\ref{int6e}) to 
(\ref{int6g}) occurs in an analog manner. Due to level repulsion,
the two states separate from one another in energy, such that their interaction
with one another is, eventually, of the same type as that with all 
the other distant states of the system. That means,  each of the original
states has ``lost'' its partner, also in this case; and the reproduction of  
the two originally  neighbored states is prevented. As a result, 
the interaction of the states of the system via the environment 
is reduced (since all states are distant in energy), with the consequence that 
the system can be described well as a closed (almost stable) system.
In difference to the case with width bifurcation, however, the number
of states of the system as a whole remains unchanged.

Eqs. (\ref{int6b}) to (\ref{int6g}) with purely imaginary and real 
coupling strength $\omega$, respectively, will seldom be realized. 
They allow us however to receive analytical results 
(see Sect. \ref{eigen}) and to understand
the basic mechanism.  Our numerical results for the more realistic
cases  with complex  $\omega$ show the same effects.
The interesting critical parameter range is that between the position 
of the EP and  that of the maximum width bifurcation or 
level repulsion, respectively,
as shown in the figures with complex $\omega$. In
this parameter range, the wavefunctions are strongly mixed and
$r_k$ varies suddenly from the value 0 at the EP,
according to  (\ref{eif5}), to the value almost 1, characteristic for
almost orthogonal states at maximum width bifurcation and level
repulsion, respectively. When $r_k \to 1$, the wavefunctions remain 
strongly mixed in relation to the original ones in all cases. 

The results are very robust and show the same characteristic
features in all cases studied by us. They hold true for systems with
loss (corresponding to decaying systems) and also for those in which gain
may occur (by absorbing particles from the environment). They hold
true also when gain and loss are balanced.

It should be underlined here once more that a quantum system is really
open and its properties are strongly influenced by the environment of
scattering wavefunctions {\it only} in the vicinity of EPs. Here,
neighboring states may strongly interact via the environment and may
cause some decoupling of the whole system from the environment, as
shown above. As a result of this decoupling, the system is stabilized; 
behaves ``linearly''; and Fermi's golden rule is applicable. 

The eigenvalues shown in Fig. \ref{fig3} left panel are real in the
parameter range between the two EPs. This might be interpreted as a
signature of $\cpt$-symmetry. The Hamiltonian $\ch^{(2)}$ is
non-Hermitian also in this parameter range. The wavefunctions are
biorthogonal and the phase rigidity is different from 1  for {\it all}
parameter values including those for which the eigenvalues are
real. The corresponding $r_i$ are near to 1, and not equal to 1. All
calculations in this parameter range for realistic systems can 
therefore be performed, to a good approximation,  by using a 
Hermitian Hamiltonian. Nevertheless,
such a calculation for an open quantum system remains an
approximation, although it will provide good results.

Moreover, the parameters used in Fig. \ref{fig3} left panel are 
unrealistic for a physical system. The coupling parameter $\omega$ is 
usually complex as discussed in, e.g., \cite{top,nearby1}. The 
eigenvalues obtained in a corresponding calculation 
are no longer real in a certain finite parameter range, see the example 
Fig. \ref{fig3} right panel. Also in this case,
PT symmetry breaking may appear and the
behavior of the system at maximum width bifurcation is determined by 
the nonlinear source term involved in the
Schr\"odinger equation for an open quantum system;
and we have the jump-like transition $r_i \to 1$ 
at maximum width bifurcation also  in this less symmetric case. 

The phenomenon of almost orthogonal wavefunctions
at maximum width bifurcation (or maximum level repulsion) is robust as 
Figs. \ref{fig1} to \ref{fig4} for different two-state systems 
show. It is {\it not} an artifact of the two-state model
(\ref{form1}) since it appears also in calculations with  more than two states
(see  Sect. \ref{num3}).

\subsection{$N>2$ states}
\label{disc3}

All our calculations with  $N>2$ states are performed in the parameter
region in which a higher-order EP is expected. Signatures of the
existence of such a higher-order EP are not found
in any of the results. This is, of course, not astonishing since every
EP is a point in the continuum and therefore of measure zero. Also the
second-order EPs (crossing points of two eigenvalue trajectories) can
be identified {\it only} by their influence onto observables in their
neighborhood. Our results show clearly that this holds true {\it also}
for higher-order EPs.

One of the characteristic features of an EP (i.e. of the crossing
point of two eigenvalue trajectories) is that the two eigenfunctions
in its surrounding are mixed due to the coupling of the system to a
common environment \cite{nearby1}. This environmentally-induced
interaction of the states is large  at and near to an EP (where the
phase rigidity of the wavefunctions is reduced, as discussed in
Sect. \ref{eigen}). A nearby state does
therefore not interact with the original states which cross at the
EP. It interacts rather with some states, the wavefunctions of which
are mixed in those of the original states (according to 
(\ref{int20})). At and near to these
crossing points, new EPs of second order appear, the ranges of
influence of which overlap. In other words, a clustering of EPs
occurs, see the discussion in section \ref{adj}.

It is this phenomenon of clustering of EPs which we 
see in  Figs. \ref{fig5} to \ref{fig7}. It
occurs in the parameter range of a higher-order EP. The results are
generic and provide us valuable information on the dynamics of open
quantum systems. Most interesting is the phenomenon of 
the jump-like enhancement of the phase rigidity when the maximum width
bifurcation is parametrically approached. This effect is of the same 
type as that observed numerically  for two states and discussed in 
detail in Sects. \ref{num2} and \ref{disc2}, respectively.

\section{Conclusions}
\label{concl}

In this paper we have described open quantum systems by means of a 
Schr\"odinger equation the Hamiltonian 
$\ch$ of which is completely non-Hermitian. It contains explicitly
(in the non-diagonal matrix elements) the interaction of the states
via the environment. The eigenvalues of $\ch$ are complex and 
the eigenfunctions are biorthogonal. 
Most interesting property is that the eigenvalues of two states may 
coalesce in one point (the so-called EP), at which also the 
corresponding eigenfunctions are the same, up to a phase \cite{top}.
The EPs are singular points and play an important role for the
dynamics of open quantum systems.

We used also the equivalent description of the system by means of a
Schr\"odinger equation with the non-Hermitian Hamilton operator
$\ch_0$ (with vanishing non-diagonal matrix elements)
and  source term. Here, the interaction of the states via the 
environment is contained in the source term, and not in the
Hamiltonian. The source term is nonlinear near and at EPs. It drives
the behavior of the open quantum system
and determines
the dynamics of open quantum systems.
    
Our main concern of the present paper is  the phase rigidity $r_i$ of
the eigenfunction $\Phi_i$ of  $\ch$. This value provides a
quantitative measure for the  biorthogonality of the wavefunction 
of the state $i$, i.e. for the possibility to influence the 
properties of the system by the environment.
It holds $1 \ge r_i \ge 0$. At $r_i \approx 1$, the wavefunctions 
are almost orthogonal, very much like  the
eigenfunctions of a Hermitian operator. For vanishing $r_i$  however, 
the eigenfunctions of the non-Hermitian operator are really biorthogonal, 
and the influence of the environment is extremely large. This influence 
may cause, among others, a mixing of the 
wavefunctions of the different states via the environment.
In \cite{nearby1,nearby2}, the modification of
the eigenfunction $\Phi_i$ of the non-Hermitian Hamilton operator $\ch$  
due to its coupling to other states of the 
system via the environment is studied
in detail. The resulting mixing 
of the wavefunctions can be expressed by the relation (\ref{int20}). 
At and near to an EP, the mixing is extremely large.

We studied first the phase rigidity in a two-level system around an
EP.  In all our calculations, the 
coupling strength $\omega$ between system and environment is
fixed. Only the energies $e_i$ or widths $\gamma_i$ of the states are
parametrically varied.  We have $r_i \approx 1 $ far from an EP 
and $r_i \to  0$ in approaching an EP.
This result is expected from analytical studies. 

We  observe however also another result in the critical region 
around an EP which is  completely unexpected. In approaching the 
maximum width bifurcation and level repulsion, respectively, the 
value of the phase rigidity varies rapidly  from its value  
$r_i < 1$ to $r_i \approx 1$. 
That means, that the two wavefunctions are almost orthogonal when the
width bifurcation or level repulsion is maximum. 
This jump-like variation of the phase rigidity is 
observed at fixed coupling strength between system and environment. 
It is caused therefore exclusively
by the nonlinear source term of the Schr\"odinger equation.
The wavefunctions of the states remain mixed at this critical
parameter value, although they are almost orthogonal according to 
$r_i \to 1$. 

This phenomenon occurs not only in the simple two-state model (see
Sect. \ref{num2})
but also in the case with more than two states (see Sect.
\ref{num3}). In the first case, we have well separated EPs while there
is some clustering of EPs in the second case. That means, the
clustering of EPs does not destroy the effect, see Figs. \ref{fig5}
to \ref{fig7}. Quite the contrary, the clustering of many EPs causes a
dynamical phase transition from an open quantum system (with
biorthogonal eigenfunctions of its states) to an almost closed system
(with almost orthogonal eigenfunctions of its states). The   
underlying process is irreversible and causes a stabilization of the
whole system, meaning that the open system 
can be described approximately as a closed system. The wavefunctions
at both sides of the dynamical phase transition 
are non-analytically related to one another and differ fundamentally 
from one another. This feature is characteristic of any phase 
transition. The wavefunctions of the states
on one side of the phase transition might be obtained by using the
two-body residual forces derived from forces between free particles.
This will be impossible, however, on the other side of the transition
where the wavefunctions are modified due to the mixing of the
different states of the system via the common environment.

Our results provide  the following generic 
feature of open quantum systems. When two states are near to one 
another in energy or in lifetime, they may strongly interact with 
one another via the continuum of scattering wavefunctions
due to the existence of a singular point (EP) in their vicinity. 
Here, the Schr\"odinger equation contains non-linear terms;  
irreversible processes occur; and the whole system will be stabilized.
As a result, the system behaves very much like a closed system
that is localized in space\,:
The eigenstates are almost orthogonal and the eigenvalues are almost 
real (and sometimes even  completely real \cite{top}). Such a 
situation can be described well by a Hermitian operator
where the lifetime of a state does not appear explicitly.
By this, the meaning of lifetime for the 
characterization of the individual states of the system is lost.
Characteristic of the states are
solely their energies and wavefunctions, while the lifetimes 
can be obtained by using perturbation methods. 

Nevertheless, the results presented in our paper show that 
energy and time are related to one another 
in quantum mechanical systems.
It is shown in [15] that time is bounded from below in
non-Hermitian quantum physics. This follows from the fact that the
decay widths (inverse proportional to the lifetimes of the states) 
cannot increase limitless. Thus, time is bounded from below in the 
same manner as energy, in contrast to the assumptions of Hermitian
quantum physics. Pauli has used this argument of Hermitian quantum 
physics, see e.g. [23], in order to conclude that the uncertainty 
relation between time and energy can, on principle, not be derived,
for details see e.g. \cite{fdp1}. The uncertainty relation 
between energy and time remained therefore a 
puzzling phenomenon in Hermitian quantum physics. As our results 
show, this phenomenon is {\it not at all} puzzling in 
non-Hermitian quantum physics.

Concluding, we recall the phenomenon of resonance trapping 
\cite{top} observed many years ago. Resonance trapping occurring 
in an open quantum system coupled strongly to the environment, 
prevents the overlapping of individual 
resonance states. Consequently, the system is practically always 
in the regime of weakly (or not) overlapping resonances, 
see e.g. \cite{top,harney}. In  analogy to this phenomenon, 
an open quantum system can be described quite well  by a 
Hermitian Hamiltonian on both sides of the dynamical phase transition. 
This statement agrees completely with experience. 
Interesting non-trivial features
of open quantum systems appear only in the parameter range in which 
a clustering of EPs and therewith a dynamical phase transition occurs.
One of many examples is the relation between reduced phase 
rigidity and enhanced transmission through a quantum dot \cite{burosa}.

The results presented in this paper are generic. We believe that 
they will initialize further studies for concrete systems under 
concrete conditions. By this, they will provide new interesting 
results for open quantum systems, especially
in the parameter range of a dynamical phase transition.


\begin{thebibliography}{99}
\bibitem{comment} In studies by some other researchers, the factor 
$i$ in (\ref{eif5}) does not appear. This difference is
discussed in detail and compared with experimental data in the
Appendix of \cite{fdp1} and in Sect. 2.5 of \cite{top}, see also
Figs. 4 and 5 in \cite{berggren}.
\end{thebibliography}
\end{document}